\documentclass[fleqn,twoside]{article}
\usepackage{espcrc2}
\usepackage{graphicx}

\begin{document}

\title{Electric Polarizability of Hadrons\thanks{Talk presented by 
       J.\ Christensen at Lattice 2002.}\thanks{This work supported 
       by DOE grant DE-FG02-95ER40907, with computer resources at 
       NERSC and JLab, and by NSF grant 0070836 and the Baylor Univ. 
       Sabbatical program, using computer resources at NCSA.}}

\author{Joe Christensen\address[McM]{Physics Department, McMurry University, Abilene, TX, 79697},
        Frank X.\ Lee\address[Jeff]{Jefferson Lab, 12000 Jefferson Avenue, Newport News,
         VA, 23606}$^{,}$\address[GWU]{Center for Nuclear Studies, George Washington University, Washington DC, 20052},
        Walter Wilcox\address[BU]{Department of Physics, Baylor University, Waco, TX, 76798},
        Leming Zhou\addressmark[GWU]}

\begin{abstract}
The electric polarizability of a hadron allows an external 
electric field to shift the hadron mass.  We try to calculate the electric 
polarizability for several hadrons from their quadratic response to the  
field at $a=0.17\,$fm  using an improved gauge field and the clover quark action.
Results are compared to experiment where available.
\end{abstract}

\maketitle

\section{Introduction}

The electric polarizability of a hadron characterizes the reaction of the quarks 
to an external electric field and can be measured by
experiment (via Compton scattering) and on the lattice.  
Conceptually, an electric field 
will tend to separate charges in a hadron, thereby affecting the internal energy of the 
hadron and thus the mass.  As in classical physics, the energy density goes as the 
square of the electric and magnetic fields:
\begin{equation}\label{eq:empol}
\Delta m = -\frac{1}{2} \, \alpha E^2 - \frac{1}{2} \,\beta B^2
\end{equation}
where ${\alpha}$ $(\beta)$ is the electric (magnetic) 
polarizability that compares to experiment.
See~\cite{Zhou:2002} for a calculation of the magnetic polarizability.

As is usual in lattice calculations, the mass of a hadron can be calculated from the 
exponential decay of a correlation function.  By calculating the ratio of the correlation 
function in the field to that without the field, we have a ratio of exponentials that 
decays as a single exponential at the rate of the mass difference.  
Equation~(\ref{eq:empol}) implies that this mass difference (between in and out of the 
field) plotted versus the electric field will be parabolic with coefficient equal to minus
half the polarizability.
By averaging $\Delta m$ over the field, $E$, and its inverse, $-E$, we hope to minimize the linear
term in the parabolic fit.

This calculation of the polarizabilities of several hadrons will follow the ideas 
discussed by Fiebig, {\it et al.\/}~\cite{Fiebig:1989en}; however, they used staggered 
fermions and we do not.  Specifically, we include the 
static E-field on the links as a phase:
(with fermion charge $q=Qe$)
\begin{equation}\label{eq:efield}
e^{iaqA} = e^{i(a^2qE)(x_4/a)} = e^{i \eta\tau} \rightarrow (1+i \eta \tau)
\end{equation}
Since the electric field is linearized in the continuum, we used the linearized 
form on the lattice.  Fiebig, {\it et al.\/} found no significant difference between the 
exponential and linearized formats for similar electric field values.
In addition, although the electric field breaks the isospin symmetry and allows 
the pion, a vector meson, to mix with glueballs, we ignore this effect in our calculation.

\section{Lattice Details}

This calculation uses the tadpole-improved clover quark action on a quenched
$12^3 \times 24$ lattice with $\beta=7.26$ ($a=0.17\,{\rm fm}$).
The gauge field was thermalized to $10^4$ sweeps and then saved every 1000.
We have 100 configurations, but only 17 were used in these preliminary results.

Including the static electric field as a phase 
on the links affects the Wilson term, but not the clover loops.
In units of $10^{-3}e^{-1}a^{-2}$, the electric field took the values $-2.16$, $4.32$, $-8.64$, and $17.28$
via the parameter $\eta=a^2QeE$ in Eq.~(\ref{eq:efield}).

With $\kappa_{\rm cr} = 0.1232(1)$, we used six values of $\kappa = 0.1182$, $0.1194$, 
$0.1201$, $0.1209$, $0.1214$, and $0.1219$, which roughly correspond to $m_q\sim 200$, 
$150$, $120$, $90$, $70$, and $50$ MeV.

\section{Results and Conclusion}

The tightness of the parabola formed by graphing the mass shift of 
a particle versus the applied electric field will give the electric polarizability
when fit with Eq.~(\ref{eq:empol}) at $B=0$.
\begin{figure}
\begin{center}
\caption{The neutron mass shift plotted versus electric field for the six 
$\kappa$ values given in the text.}
\label{f:mass-shift}
\includegraphics[width=5.25cm, angle=90]{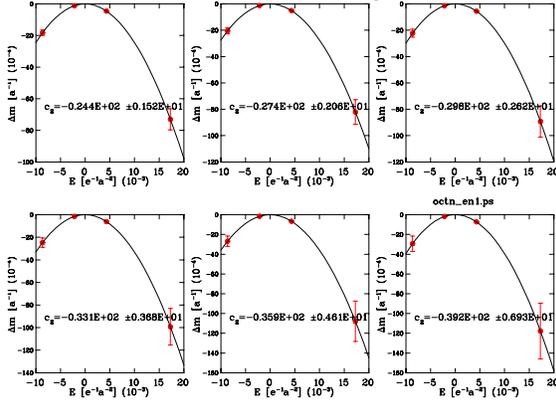}
\end{center}
\end{figure}
The results for the neutron are shown in Fig.~\ref{f:mass-shift}.
The graphs are sorted by mass, large to small displayed from the 
upper-left across and down to the lower-right.  
The nice parabolic fit of Fig.~\ref{f:mass-shift} indicates that 
averaging over both directions of the field, $E$ and $-E$, has 
minimized any linear dependence.
Similar graphs can be produced for the $\Delta^0$, $\pi^0$ and $\rho$
to varying degrees of confidence.  

\begin{figure}
\begin{center}
\caption{The polarizability of the neutron plotted versus $m_\pi^2$ as
determined from the six graphs of Fig.~\protect{\ref{f:mass-shift}}. }
\label{f:pol}
\includegraphics[width=5.25cm, angle=90]{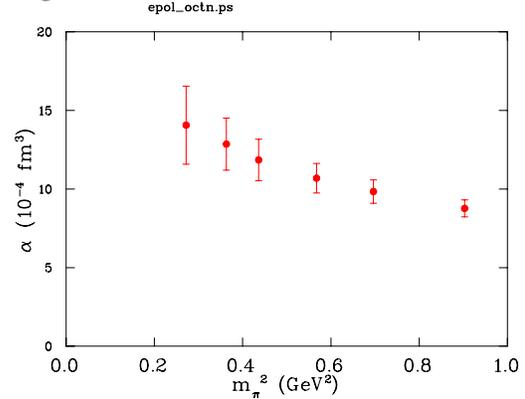}
\end{center}
\end{figure}

With more configurations, we will consider the chiral limit by 
combining the quadratic coefficient from these results into a plot
of polarizability versus mass.  Fig.~\ref{f:pol} shows the 
neutron polarizability increasing in this limit.
With only 17 configurations, our preliminary result is not quite 
consistent with the average of the tabulate results, $10.4 \times 10^{-4}\,{\rm fm^3}$.

\begin{figure}
\begin{center}
\caption{The polarizability of the $\pi^0$ plotted versus $m_\pi^2$.}
\label{f:pi-pol}
\includegraphics[width=5.25cm, angle=90]{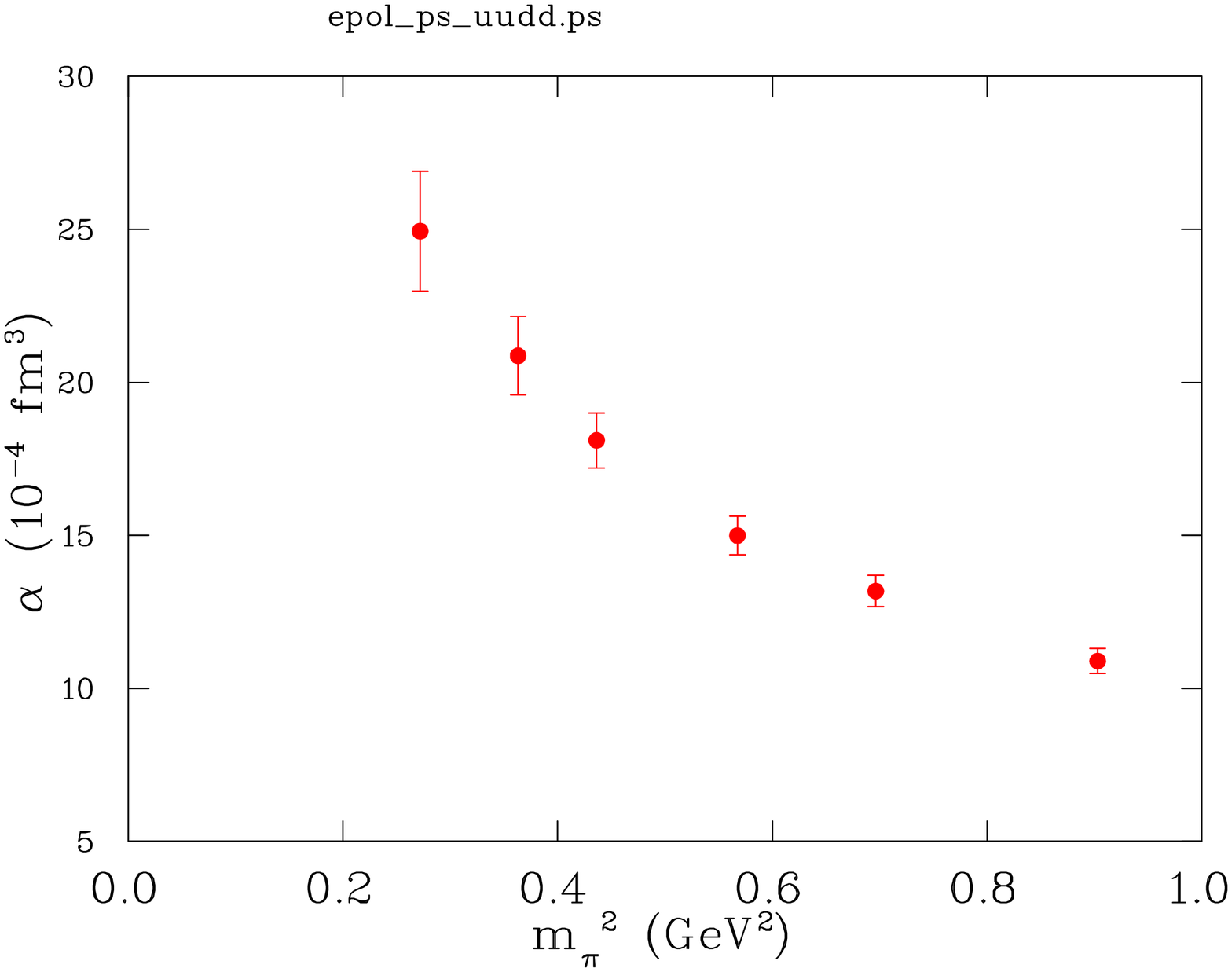}
%
%
\vspace{-1cm}
\caption{The polarizability of the $\Delta^0$ plotted versus $m_\pi^2$.}
\label{f:del-pol}
\includegraphics[width=5.25cm, angle=90]{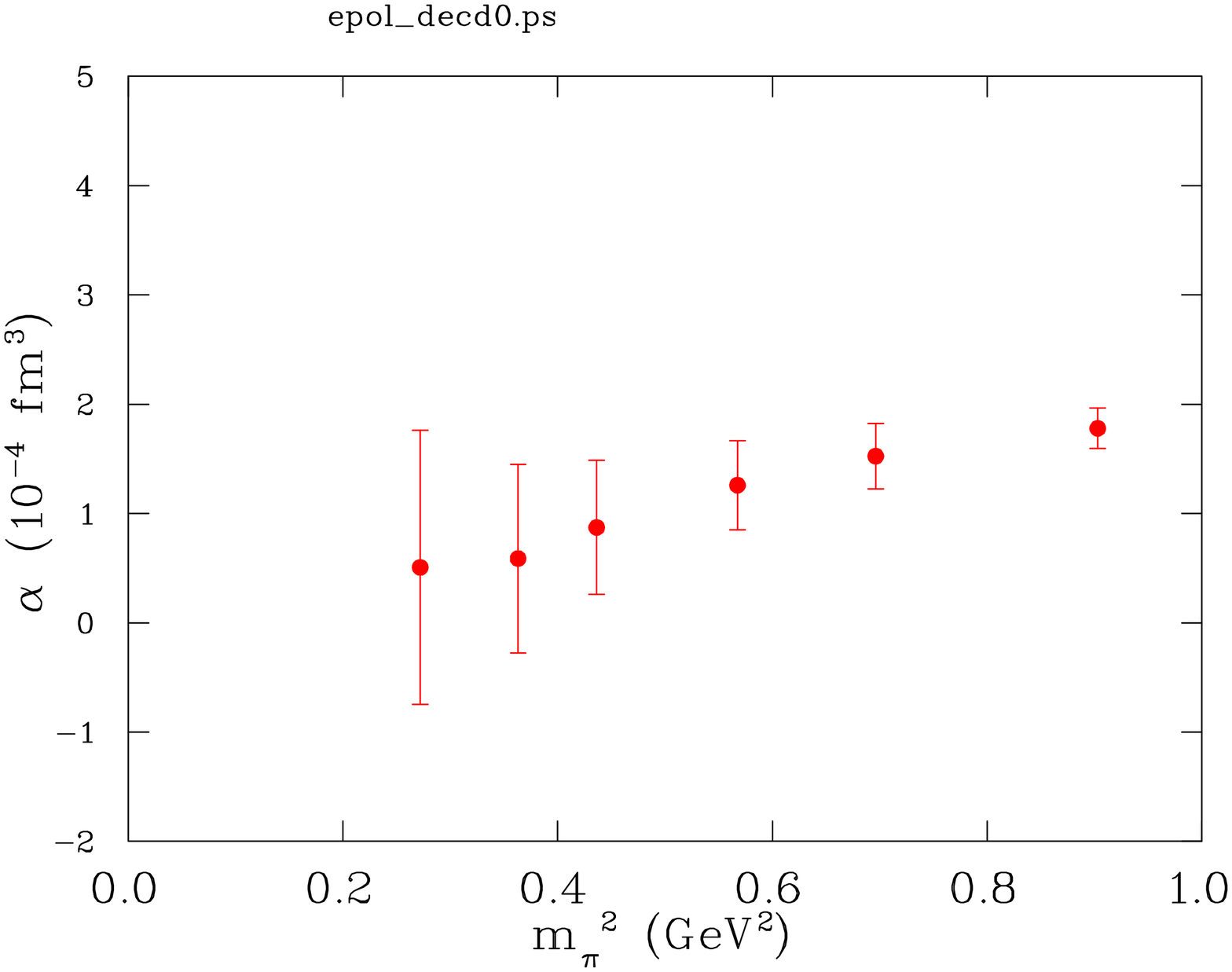}
%
%
\vspace{-1cm}
\caption{The polarizability of the $\rho^0$ plotted versus $m_\pi^2$.}
\label{f:rho-pol}
\includegraphics[width=5.25cm, angle=90]{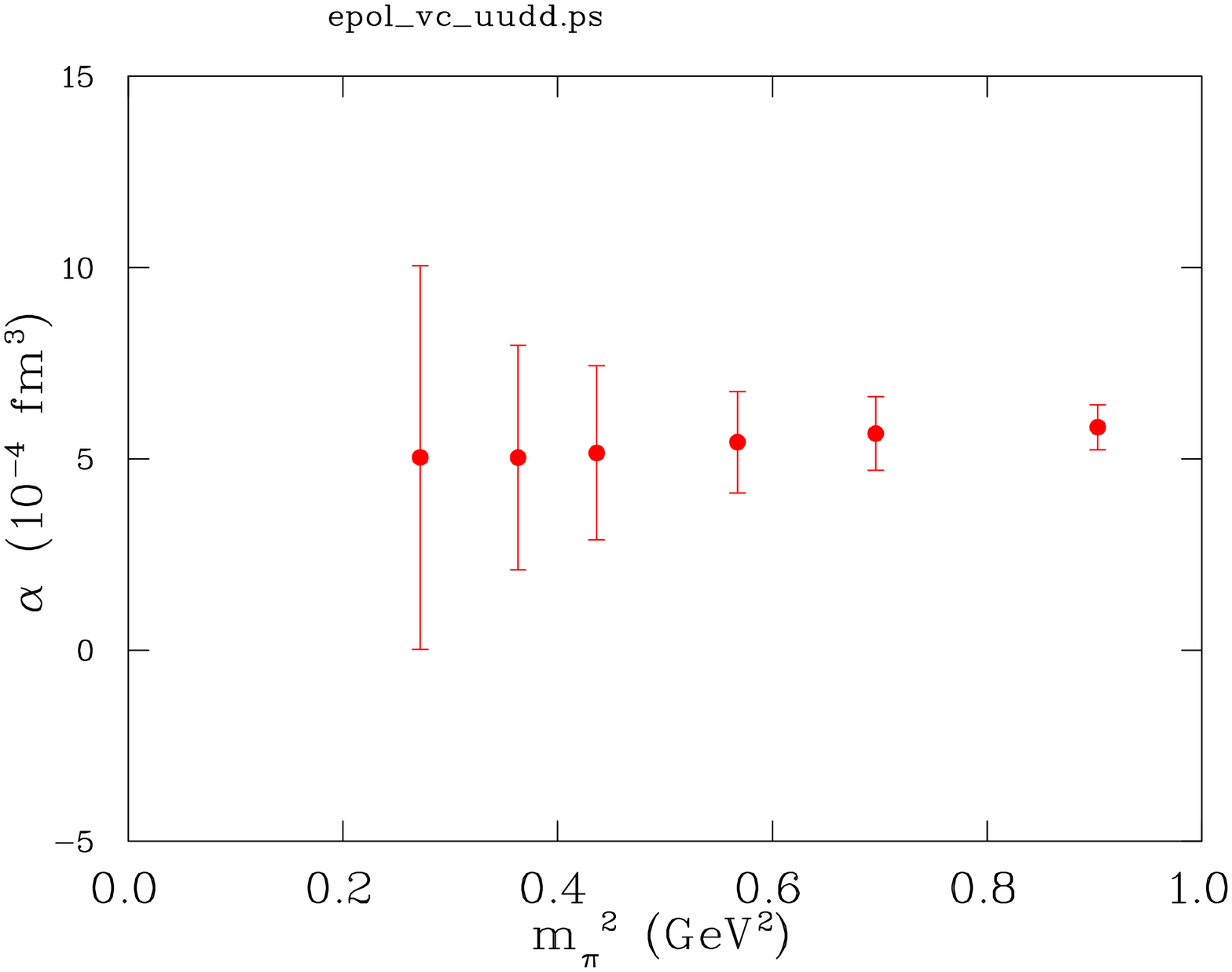}
\end{center}
\end{figure}

Although $\chi$PT~\cite{Portoles:1994rc} predicts that $\alpha_\pi$ will be 
small and negative, $\approx -0.5\times 10^{-4}\,{\rm fm^3}$,
Fig.~\ref{f:pi-pol} shows peculiarly large results.
The delta polarization seems to be dropping off in Fig.~\ref{f:del-pol},
and amidst large error-bars, the rho of Fig.~\ref{f:rho-pol} seems quite flat.

Table~\ref{t:world} lists the following subset of existing 
values for the neutron polarizability.
There is by no means a consensus in the results.
\begin{table}
\begin{center}
\caption{Current experimental values (\cite{Rose:1990:1990eu,Schmiedmayer:1991,Koester:1995nx,Kolb:2000ix}
with modifications by \cite{Lvov:1993fp,Enik:1997cw})
and theoretical values (back to 2000) for the polarizability of the 
neutron.  
Several calculations are refinements to previous calculations.
$\alpha_n$ is listed in units of $10^{-4}\,{\rm fm^3}$.}
\label{t:world}
\begin{tabular}{ll}
ref: method & $\alpha_{\rm n}$ \\ \hline
\cite{Rose:1990:1990eu}: Exp: Quasi-Free $\gamma d$ & $10.7^{+3.3}_{-10.7}$ \\
\cite{Schmiedmayer:1991,Lvov:1993fp}: Exp: $n$-$^{208}$Pb & $12.6\pm1.5\pm2.0$ \\
\cite{Koester:1995nx,Lvov:1993fp}: Exp: $n$-$^{208}$Pb & $0.6\pm5$ \\
\cite{Kolb:2000ix}: Exp: uses \cite{Rose:1990:1990eu} & $13.6^{+0.4}_{-6.0}$ \\
\cite{Enik:1997cw}: claims \cite{Schmiedmayer:1991,Lvov:1993fp} is & $7-19$ \\
\cite{Levchuk:1999zy}: Potential model & $9\pm3$ \\
\cite{Kondratyuk:2001qu}: Dressed K model $\!\!\!\!\!\!$ & $12.7$ \\
\cite{Kossert:2002jc}: using \cite{Levchuk:1999zy}'s results & $12.5\pm1.8^{+1.1}_{-0.6}\pm1.1$ \\
\cite{Lundin:2002jy}: with \cite{Levchuk:1999zy}'s authors & $9.2\pm2.2$
\end{tabular}
\end{center}
\end{table}
Experimentally, $\alpha_n$ is found through Compton scattering off of a deuteron
and the polarizability of a proton must be included in the model.  In this data, 
there are four experimental results,~\cite{Rose:1990:1990eu,Schmiedmayer:1991,Koester:1995nx,Kolb:2000ix};
but \cite{Kolb:2000ix} incorporates \cite{Rose:1990:1990eu}.
L'vov~\cite{Lvov:1993fp} corrected the earlier values by adding $0.6\times 10^{-4}\,{\rm fm^3}$ to their results.
The experiments measure $\alpha-\beta$ and the results are found using 
a sum rule that gives $\alpha+\beta$.
The results are quite model-dependent, and even 
the experiments do not agree.
We note that \cite{Enik:1997cw} claims
\cite{Schmiedmayer:1991,Lvov:1993fp} miscalculated their errors.
Further, \cite{Kossert:2002jc} used \cite{Levchuk:1999zy}'s results to get a different number
and then the authors of \cite{Levchuk:1999zy} joined others \cite{Lundin:2002jy} to find a result close to 
their original value.

With consistent results for the neutron, we believe that finishing this calculation 
on the rest of our configurations will give results that can reliably be compared 
to the experimental results.


\end{document}